\documentclass{aa}
\usepackage{psfig}
\def\earth{\hbox{$\oplus$}}
\newcommand{\et}{{\it et al.}}

\begin{document}

   \thesaurus{} 
   
   \title{Intra-Day Variability and the Interstellar Medium Towards 0917+624}

   \author{D.L. Jauncey\inst{1} \& J.-P. Macquart\inst{2}}

%   \offprints{D. Jauncey}

   \institute{$^1$Australia Telescope National Facility,  CSIRO, 
   Epping, NSW 2121, Australia, email: djauncey@atnf.csiro.au\\
	      $^2$Research Centre for Theoretical Astrophysics, School 
	     of Physics, University of Sydney, NSW 2006, Australia}

   \date{Received December x, 2000; accepted }

   \authorrunning{Jauncey \& Macquart}
   \titlerunning{Annual variability in the IDV of 0917$+$624}
   \maketitle

\begin{abstract}
The intra-day variable source 0917$+$624 displays annual changes in 
its timescale of variability.  This is explained in terms of 
a scintillation model in which changes in the variability timescale are due to 
changes in the relative velocity of scintillation pattern as the Earth orbits
the sun.

\end{abstract}

\section{Introduction}
The radio source S5 0917+624, identified with a $z = 1.446$ quasar
(Stickel \& Kuhr 1993), was one of the first discovered to show
strong intra-day variability (IDV) at radio wavelengths (Heeschen
\et\ 1987). Recently, Kraus \et\ (1999) reported three-epoch
5 GHz observations, December 1997, September 1998 and
February 1999, which show significant changes in the IDV
characteristics of this source. They suggest several possible
interpretations which have at their basis structural changes, and
hence brightness temperature changes, within the source itself.

Alternatively we suggest that the observed changes in IDV result
from variations in the interstellar medium (ISM) velocity as the
Earth revolves around the sun. We present
evidence to support this suggestion
based on published data on the IDV of 0917+624.

\section{Changes in the Intra-Day Variability in 0917+624}

Kraus \et\ (1999) undertook to explain the observed changes in the IDV 
properties as an intrinsic change in the source.  This is irrespective 
of whether the IDV results from intrinsic variability or from 
interstellar scintillation (ISS).  If the variability is intrinsic, 
changes in the IDV timescale imply source size changes directly, while 
if the variability is due to ISS, it was argued that the change in 
timescale could be either the due to disappearance of the compact 
component or an increase in its angular size sufficient to quench the 
scintillation.  Here we offer an alternative explanation based on an 
ISS origin of the variability, showing that changes in the timescale 
need not reflect changes in source structure.

Over the course of a year the motion of the Earth, projected
on the plane of the sky is an ellipse with eccentricity dependent on the
ecliptic latitude of the source. Extragalactic sources show negligible proper
motion, and hence the relative velocity of the ISM across
the line of sight to the source is simply the difference between
the ISM and the Earth's motion. If these
are comparable in amplitude, then there may be a period
when the vectors are parallel, and six months later, anti-parallel. In the 
first case the apparent ISM velocity as
seen from the Earth is low, in the latter, high.

For an ISS origin of IDV, it follows that at times of low
relative velocity, the characteristic time-scale of the
variability will be lengthened, while six months displaced,
it will be correspondingly shortened. This form of behaviour
has been seen in the rapidly variable IDV source
J1819+3845 (Dennett-Thorpe \& de Bruyn, 2000), and
strongly supports the ISS origin for the IDV in this source.

We have searched the literature on flux density
monitoring of 0917+624 to determine its behaviour. The first observations
were made in 1985 (Witzel \et\ 1986; Heeschen \et\ 1987) at 2.7 GHz 
with the Bonn 100~m telescope. Since then we have located a total of 10 well
documented observing sessions undertaken with either the
Bonn telescope or the VLA, and these are listed in Table~1.

From this data we have determined the characteristic time scale, 
$T_{\rm char}$, (defined for convenience, as the mean time between 
successive peak-to-trough or trough-to-peak excursions) for each 
session at 2.7 and/or 5 GHz.  This can only be done when the session 
was of sufficient length that several peaks/troughs are present in the 
data.  Our choice of $T_{\rm char}$ corresponds to the timescale on 
which the intensity structure function saturates, and hence closely 
follows the definition used by Kraus \et\ (1999).  Table~1 shows that 
$T_{\rm char}$ is close to one day at 2.7~GHz, and roughly half this 
at 5~GHz.

Table~1 also shows that the observing sessions for 0917$+$624
are not uniformly spread throughout the year, most are
concentrated around the northern winter and spring.  The only other 
session that we found that shows a slowing down of the IDV similar 
to that observed in September 1998 was the August 1985 session at 
2.7~GHz (Heeschen \et\ 1987).  These two observations 13 years apart 
clearly establish an annual cycle in this source.  

\section{Effect of the Earth's orbital motion}
Changes in the variability timescale of a scintillating source 
occur due to changes in both the direction and speed of the scintillation 
pattern as it moves across the source-observer line of sight.  The 
scintillation timescale is 
\begin{eqnarray}
t_{\rm ISS} = s_0 (\alpha)/v_{\rm app}, \label{tISS}
\end{eqnarray}
where $s_0(\alpha)$ is the length scale of the scintillation pattern in 
the direction parallel to the apparent scintillation velocity, 
$\alpha$ is the direction of motion relative to the pattern, and 
$v_{\rm app}$ is the apparent speed of the scintillation pattern 
relative to the Earth.  The scintillation timescale is formally 
defined as the timescale on which the autocorrelation function of 
intensity fluctuations reaches 1/e of its saturation value; operationally, 
$t_{\rm ISS} \approx 0.6\,T_{\rm char}$.
We assume that the scattering occurs only at a single 
screen, so that the intrinsic velocity of the scattering 
medium relative to the heliocentre is described by one velocity.  

\subsection{Changes in the scintillation speed}
Annual changes in the scintillation speed result in an annual modulation of 
the variability timescale.  The apparent scintillation speed 
is the magnitude of the component of the ISM's intrinsic velocity, 
relative to the Earth's velocity, that is tangential to the line of sight. 
When the Earth's velocity nearly matches the 
intrinsic velocity of the scattering material, the scintillation 
pattern would then appear to be stationary, causing the variability to stop.

%
% NB: see http://nssdc.gsfc.nasa.gov/planetary/factsheet/earthfact.html
%
For a source at ecliptic latitude $\theta$ and longitude $\phi$, it is 
convenient to denote the Earth's velocity as
\begin{eqnarray}
{\bf v}_{\earth}= v_{\earth} (-\sin \Psi, \cos \Psi,0),
\end{eqnarray}
where $v_{\earth} = 29.8$~km/s is the Earth's orbital speed about the
heliocentre and the orbital phase is  
$\Psi=2 \pi t - \phi$, and $t$ is the time 
in years measured from the vernal equinox.  The apparent scintillation 
speed is then
\begin{eqnarray}
v_{\rm app} = [(v_{\parallel} + v_{\earth} \cos \Psi)^2 + 
(v_{\perp} + v_{\earth} \sin \theta \sin \Psi)^2 ]^{1/2}, 
\label{ScintSpeed}
\end{eqnarray}
%
% nb: check sign of vperp/vparallel +/- vearth sin/cos Psi. : 
% original equn has +!  It all
% depends on the direction of rotation (clockwise, counterclockwise)
%
where $v_{\parallel}$ is defined as the speed of the 
scattering screen parallel to the Earth's ecliptic at closest 
approach, which occurs at $\Psi=\phi/2 \pi$, 
and $v_{\perp}$ is its speed perpendicular to the ecliptic at this 
point, with positive values of $v_{\perp}$ indicating motion toward 
higher ecliptic latitudes.

\subsection{Changes in the scintillation direction}

Changes in the direction of the scintillation pattern, $\alpha$, influence the 
scintillation timescale only if the scintillation pattern is not circularly 
symmetric, i.e. if $s_0$ depends on $\alpha$.  
Asymmetry in the scintillation pattern may be due to either intrinsic source 
structure or anisotropy in the scattering medium.  

The effect of source structure is important when the angular size of 
the source exceeds the angle to which a point source would be 
scatter-broadened by the scintillation pattern, $\theta_{\rm pt}$.
The scintillation lengthscale, $s_0$, for a source of angular size 
$\theta_S(\alpha)$ is (e.g. Narayan 1992)
\begin{eqnarray}
s_0(\alpha) \approx \left\{ \begin{array}{ll}
D \, \theta_{\rm pt}, & \theta_S(\alpha) < \theta_{\rm pt} \\
D \, \theta_S(\alpha), & \theta_S(\alpha) > \theta_{\rm pt}
\end{array}, \right. \label{scintscale} 
\end{eqnarray}
where $D$ is the distance to the scattering screen.  

The angular scale, $\theta_{\rm pt}$, depends on the strength 
of the scattering, which is a function of frequency.  Above the
transition frequency, $\nu_0$, the scattering is weak, and 
$\theta_{\rm pt}$ corresponds to the angular scale of the first 
Fresnel zone, $\theta_{\rm F} = \sqrt{c/(2 \pi \nu D)}$.  

In the regime of strong scattering at frequencies below $\nu_0$, one 
has $\theta_{\rm pt} = \sqrt{c/(2 \pi \nu_0 D)} (\nu/\nu_0)^{-2.2}$.  The 
exponent $2.2$ follows from the assumption that the turbulence in the 
scattering medium follows a Kolmogorov scaling 
(e.g. Armstrong, Rickett \& Spangler 1995).  The scale of size of the 
scintillation pattern of a point-like source increases strongly towards 
lower frequencies.

IDV in extragalactic radio sources straddles
the frequency range where weak and strong scattering occurs.
This is particularly important in the case of 0917$+$624.  Weak 
scattering is likely to be applicable to observations of 
0917$+$624 at frequencies $\nu \ga 5$~GHz (Walker 1998), while the 
timescale of variability observed at $2.7$~GHz is likely to 
correspond to refractive scintillation in the transition to the 
strong scattering regime.

% define anisotropy
Anisotropy in the scattering medium causes elongation of the 
scintillation pattern in a direction orthogonal to the magnetic field 
in the scattering medium (e.g. Goldreich \& Sridhar 1995).  
Observations suggest elongations up to $\sim 2:1$ are possible 
(e.g. Wilkinson, Narayan \& Spencer 1994; Spangler \& Cordes 1998).
The effect of anisotropy has been discussed by Backer \& Chandran 
(submitted) on the scintillation parameters of nearby pulsars and 
IDV radio sources.  Effects 
due to anisotropy become unimportant for source sizes larger than 
$\theta_{\rm pt}$, since the scale of the scintillation pattern is 
then dominated by the angular size of the source itself.

\subsection{Application to 0917$+$624}\label{0917model}
We show that the changes in the scintillation speed 
due to variations in the Earth's orbital velocity alone are sufficient 
to reproduce the behaviour of 0917$+$624.  We consider only changes in the 
scintillation speed, and neglect any possible asymmetry in the source 
structure.  We also neglect the effect of anisotropy in the scattering medium 
because the characteristic timescale of the intraday variability in 0917$+$624 is much 
longer than the Fresnel timescale, suggesting an angular size 
$\theta_S(\alpha) \ga \theta_{\rm F}$.  (Numerically, one has 
$\theta_F=2/4 \times 10^8$~m for a screen of $D=200$~pc at $\nu=5$~GHz.)

The ecliptic latitude and longitude of this source are $\theta=0.768$ 
and $\phi = 2.08$ respectively.  We find ISM velocities of 
$v_{\perp}=-16 \pm 3$~km/s and $v_{\parallel} = -16 \pm 3$~km/s 
reproduce the sharpness of peaks observed $T_{\rm char}$ at 2.7 and 
5~GHz.  The scale size of the scintillation pattern, $s_0$, may be 
calibrated by a measurement of the variability timescale given 
an estimate of the scintillation speed at that time.  For the 
$v_{\perp}$ and $v_{\parallel}$ required to reproduce the observed 
data this implies $s_0 \approx 9.5 \times 10^8$~m at 5~GHz and $s_0 
\approx 1.9 \times 10^{9}$~m at 2.7~GHz.  For a scattering screen at 
$D=200$~pc, as used by Rickett \et\ (1995), this corresponds to 
angular sizes $\theta_S = 30$ and $65\,\mu$as at 5 and 2.7~GHz 
respectively.  This is quite consistent with the derived angular sizes of 
other IDV sources (e.g. Macquart \et\ 2000).

The comparison between the observational data and the model is shown 
in Figure 1.  The model gives excellent agreement with the data at 
both 5 and 2.7~GHz.  In particular, it clearly reproduces the peak in 
the August and September values of $T_{\rm char}$.
The location of the peak observed in Fig. 1 is strongly sensitive to 
the ecliptic longitude of the source, but only weakly dependent on 
the intrinsic velocity of the ISM.

% Jauncey's earlier bit moved here...mesh with my stuff
 
% At 5 GHz where there is the most data, the
% September 1998 point is at least 9 times larger than the
% mean of the other 7 measurements. 

The validity of the above model can be readily checked with 
more data at both frequencies over the July through 
November period.  Our model predicts a 
steady increase in the characteristic time scale during July, a 
significant slow-down in the variability through August and into September, with a 
decreasing time scale probably October through November.

\section{Discussion}

\subsection{The intrinsic brightness temperature of the source}
Examination of the light curves during the slow-down period
can also be used to set limits on any intrinsic variability that
may be present at radio wavelengths. During the slow-down
the scintillations are virtually suppressed, so that any
residual variations observed provide an upper limit to the
intrinsic variations. This is an upper limit because there are
probably low-level scintillations remaining as the model of a
simple screen at a single distance with a single velocity is
certainly an oversimplification.

For 0917+624, the August and September data still show the
presence of low-level variability. Because of the $\sim 5$ day
extent of both sessions, the derived variability brightness
temperature, estimated using a source diameter of 5 light days, 
is $T_{vb} < 10^{16}\,$K, still well above the inverse
Compton limit, but significantly less than the $10^{18}\,$K inferred
if the intra-day variability were intrinsic.  Future observations
of longer duration may better define the slow-down
period and further reduce this limit.

\subsection{Earth Revolution Scintillation Synthesis}

The annual procession of the Earth around the sun and its
effects on ISS shows a number of parallels with Earth
rotation synthesis: in the latter we have a regularly
constructed interferometer that completes its synthesis
in half a day (half of the Earth's rotation). In the former case,
we have an irregular ``interferometer'', namely the ISM,
and it completes its ``synthesis'' in half a year, which corresponds to 
half of the Earth's rotation around the sun. 

We suggest that Earth Revolution Scintillation Synthesis
can be used to model the
microarcsecond structure of compact radio sources. Once the ISM velocity has been
determined for a particular source, the the shape of the
scintillation light curves may provide information on
source structure. 

The technique can be extended to studying the polarization
structure. Macquart \et\ (2000) make 
use of this in studying the
linearly and circularly polarized light curves for the strong
and variable linear and circular polarization in the southern
IDV source PKS~1519$-$273. Here both the linear and circular
light curves follow the total intensity light curve quite closely,
and it is clear that the microarcsecond source that is
responsible for the IDV has the same structure in the
total intensity, and linearly and circularly polarized components.

\section{Conclusion}
We have shown that a simple scintillation model accounts for the 
annual changes in the intraday variability timescale observed in the source 
0917$+$624.  These changes are due to the Earth's orbital motion, 
which influences the variability timescale because its velocity with 
respect to the ISM changes with time.  This causes a 
change in the speed at which the Earth moves across the scintillation 
pattern.

This strongly suggests that ISS is the dominant mechanism of 
intra-day variability in this source.  

We urge further observations of this source in the period July to 
November to test this model and constrain the velocity of the ISM 
transverse to the line of sight towards 0917$+$624.

We also suggest that observations of the source's variability properties 
at various epochs can be used to gain information on the two-dimensional 
structure in the  source, as the direction of motion across the scintillation 
pattern varies during the year.

\medskip
\noindent
{\bf Final Note}
During final preparation of this paper, we heard that Rickett, Witzel, 
Kraus, Krichbaum \& Qian have independently discovered the annual cycle in
0917$+$624 and the explanation in terms of ISS  
and are submitting an independent publication.

\begin{acknowledgements}
We thank Lucyna Kedziora--Chudczer, Jim Lovell and Don Melrose for 
valuable discussions.  The ATNF is funded by the Commonwealth 
Government for operation as national facility by CSIRO.
\end{acknowledgements}

% \vspace{0cm}
% \hbox{\hspace{-7cm}\psfig{file=5GHznew.eps, width=148mm}\hspace{0cm}
% \psfig{file=2GHznew.eps, width=148mm}}
% \vspace{0cm}

% \hfill\parbox[b]{5.5cm}{\caption[a]{Comparison of the data at (a) 
% 5~GHz and (b) 2~GHz with the scintillation model described in 
% \S\ref{0917model} with $v_{\parallel} = 20$~km/s, $v_{\perp}= 15$~km/s.
% Scintillation pattern sizes of 
% $s_0 = 2.0 \times 10^$~m and $s_0=4.0 \times 10^6$~m have 
% been used at 5 and 2.7~GHz respectively (see text).}}

   \begin{table}
      \caption[]{Values of $T_{\rm char}$ measured from the literature.}
         \label{MeasuredTchars}
      \[
         \begin{tabular}{ccccl}
            \hline
            \noalign{\smallskip}
	    Mean Date &  $\nu$ (GHz) & modulation & $T_{\rm char}$ & 
	    Reference  \\
            (year) & & index (\%) & (days) &  \\
	    \hline
	    \noalign{\smallskip}
  1985.63 & 2.7 & 0.9     & $>4$ &  a  \\
  1985.99 & 2.7 & $>2.4$  & $1.3 \pm 0.1$ & a \\
  1988.26 & 2.7 & 5.7    & $0.9 \pm 0.1$ & b \\
  1988.46 & 2.7 & 4.6    & $1.1 \pm 0.1$ & b\\
  1989.00 & 2.7 & 7.3    & $1.2 \pm 0.1$ & c \\
  1989.35 & 2.7 & 8.3    & $0.7 \pm 0.1$ & d \\
  \noalign{\smallskip}
  1988.26 & 5 &   3.7    & $0.3 \pm 0.1$ & b \\
  1988.46 & 5 &   3.1    & $0.7 \pm 0.1$ & b \\
  1989.00 & 5 &   6.0    & $0.6 \pm 0.1$ & c \\
  1989.35 & 5 &   4.4    & $0.4 \pm 0.1$ & d\\
  1990.12 & 5 &   3.7    & $0.6 \pm 0.1$ & d \\
  1997.98 & 5 &   5.2    & $0.7 \pm 0.1$  & e\\
  1998.71 & 5 &   1.8    & $>5$  & e\\
  1999.10 & 5 &   5.1    & $0.8 \pm 0.3$ & e\\
            \noalign{\smallskip}
	    \hline
         \end{tabular}
      \]
\begin{list}{}{}
\item[a] Heeschen et al. 1987, {\it A.J.}, {\bf 94}, 1493
\item[b] Quirrenbach et al. 1989a,  {\it Nature}, {\bf 337}, 442
\item[c] Quirrenbach et al. 1989b, {\it A.\&A.}, {\bf 226}, L1
\item[d] Quirrenbach et al. 2000, {\it  A.\&A. Supp. Ser.}, {\bf 141}, 221
\item[e] Kraus et al. 1999, {\it A.\&A.}, {\bf 352}, L107
\end{list}
   \end{table}

% Table 1
% 
% Tchar defined as the mean time between a maximum and the next minimum
% (or a minimum and the next maximum).
% 
% Mean Date		freq	modI	Tchar	Reference
% 		GHz	\%	days
% 
% 1985.63	Aug	2.7	0.9 \%	>4	Heeschen et al., 1987
% 
% 1985.99	Dec	2.7	>2.4 \%	1.3	Heeschen et al., 1987
% 
% 1988.26	Mar/Apr	2.7	5.7 \%	0.9	Quirrenbach et al., 1989a
% 
% 1988.46	June	2.7	4.6 \%	1.1	Quirrenbach et al., 1989a
% 
% 1989.00	Dec/Jan	2.7	7.3 \%	1.2	Quirrenbach et al., 1989b
% 
% 
% 
% 1988.26	Mar/Apr	5	3.7 \%	0.3	Quirrenbach et al., 1989a
% 
% 1988.46	June	5	3.1\%	0.7	Quirrenbach et al., 1989a
% 
% 1989.00	Dec/Jan	5	6.0 \%	0.6	Quirrenbach et al., 1989b
% 
% 1989.35	May	5	4.4 \%	0.4	Quirrenbach et al., 2000
% 
% 1990.12	Feb	5	3.7 \%	0.6	Quirrenbach et al., 2000
% 
% 1997.98	Dec	5	5.2 \%	0.7 (0.9)	Kraus et al., 2000
% 
% 1998.71	Sept	5	1.8 \%	> 5	Kraus et al., 2000
% 
% 1999.10	Feb	5	5.1 \%	0.6 (1.3) 	Kraus et al., 2000
% 
% 
% References used to construct the above data set for 0917+624 IDV
% 
% Heeschen et al., (1987)  AJ., vol 94, p 1493
% 
% Quirrenbach et al., (1989a)  Nature, vol 337, p 442, 1989
% 
% Quirrenbach et al., (1989b)  A\&A, vol 226, L1
% 
% Quirrenbach et al., (1992)  A\&A, vol 258, p 279
% 
% Kraus et al., (1999)  A\&A, vol 352, p L107
% 
% Quirrenbach et al., (2000)  A\&A Supp Ser, vol 141, p 221

\vspace{0.3cm}
\hbox{\hspace{-8cm}\psfig{figure=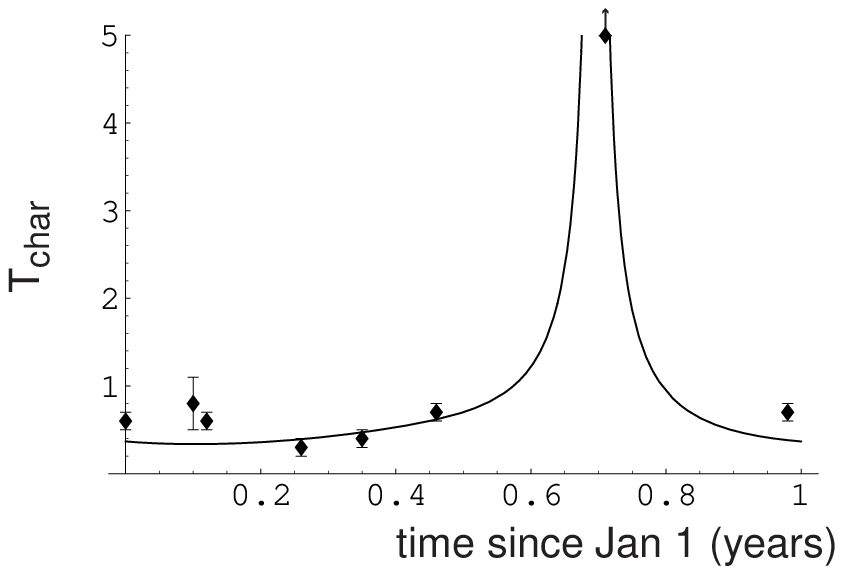,width=7.4cm}\hspace{0cm}
\psfig{figure=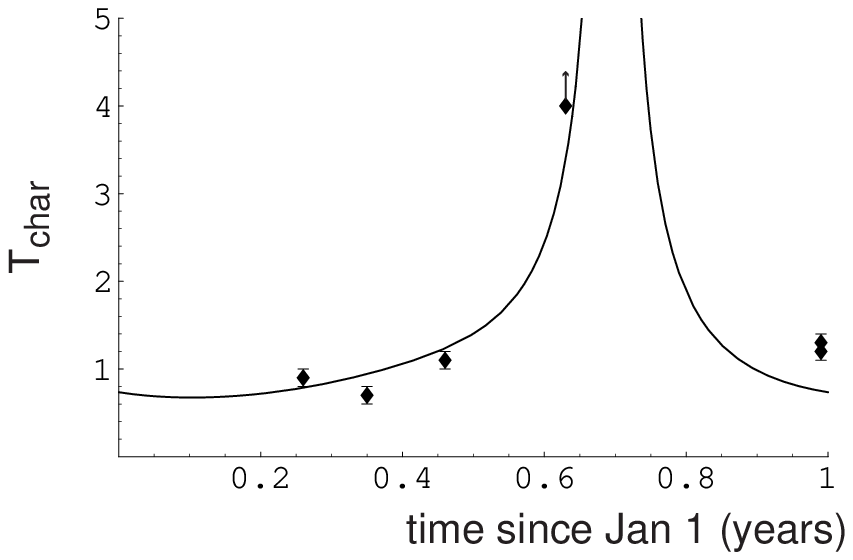,width=7.4cm}}
\vspace{0cm}

{\bf Figure 1}: Comparison of the data at (a) 5~GHz and (b) 2~GHz with the 
scintillation model described in \S\ref{0917model} with $v_{\parallel} 
= -18$~km/s, $v_{\perp}= -16$~km/s.  Scintillation pattern sizes of $s_0 
= 9.5 \times 10^8$~m and $s_0=1.9 \times 10^9$~m have been used at 5 
and 2.7~GHz respectively (see text).

\end{document}